\documentstyle[preprint,tighten,pra,aps,epsf,psfrag]{revtex}
\begin{document}
\tightenlines
\draft

\preprint{UTF 408}

\title{Rotating topological black holes}

\author{D.~Klemm\thanks{email: klemm@science.unitn.it}}
\address{Dipartimento di Fisica, Universit\`a  di Trento, Italia}
\author{V.~Moretti\thanks{email: moretti@science.unitn.it}}
\address{ECT* European Centre for Theoretical Studies\\ 
in Nuclear Physics and Related Areas,\\ 
Trento, Italia\\ 
and Istituto Nazionale di Fisica Nucleare,\\ 
Gruppo Collegato di Trento, Italia} 
\author{L.~Vanzo\thanks{email: vanzo@science.unitn.it}}
\address{Dipartimento di Fisica, Universit\`a di Trento 
\\ and Istituto Nazionale di Fisica Nucleare 
\\ Gruppo Collegato di Trento, Italia}

\maketitle\begin{abstract}
A class of metrics solving Einstein's equations with negative 
cosmological constant and representing rotating, topological black 
holes is presented. All such solutions are in the Petrov type-$D$ 
class, and can be obtained from the most general metric known in this 
class by acting with suitably chosen discrete groups of isometries. 
First, by analytical continuation of the Kerr-de 
Sitter metric, a solution describing uncharged, rotating black holes 
whose event horizon is a
Riemann surface of arbitrary genus $g > 1$, is obtained. Then a 
solution representing a rotating, uncharged toroidal black hole is also 
presented. The higher genus black holes appear to be quite exotic 
objects, they lack global axial symmetry 
and have an intricate causal structure. The toroidal black 
holes appear to be simpler, they have rotational symmetry and the 
amount of rotation they can have is bounded by some power of the mass. 
\end{abstract}


\maketitle

\section{Introduction}

In the past months there has been an increasing interest in black holes
whose event horizons have a nontrivial topology \cite{amin,mann97,vanz97}.  
The solutions can be obtained with the least expensive modification 
of general relativity, the introduction of a negative cosmological 
constant. This is sufficient to avoid few classic theorems 
forbidding non-spherical black holes\cite{gann76,hawk72,frie93}, and comes 
as a happy surprise. Charged versions of these black holes were presented in 
\cite{mann97}, they can form by gravitational 
collapse\cite{manW97,lemo97} of 
certain matter configurations, and all together form a 
sequence of thermodynamically well behaved objects, 
obeying the well known entropy-area law \cite{vanz97,bril97}.\\ 
Up to now no rotating generalization of higher genus solutions has been known. 
Holst and Peldan recently showed that there does not exist any (3+1)-
dimensional generalization of the rotating BTZ black hole \cite{holst97}.
Therefore, if we are looking for a rotating generalization of the
topological black holes, we have to consider spacetimes with non constant
curvature.\\ 
On the other hand, a charged rotating toroidal solution with a black 
hole interpretation has been presented by Lemos and Zanchin 
\cite{lemzan}, following previous work on cylindrically symmetric 
solutions of Einstein's equations 
\cite{levi,chaz,curz,lewi,linet,santo,lemocy}.\\ 
In this paper a rotating generalization of higher genus black holes 
together with another toroidal rotating solution will be 
presented. We do not present uniqueness results, and apart from noticing 
that there are more than one non isometric tori generating black 
holes, we content ourselves 
with a discussion of some of the relevant properties they have.\\ 
We begin in Sec.(II) with the spacetime metric for the genus $g>1$ 
solution and give a proof that it solves Einstein's field equations 
with negative $\Lambda$-term.\\ 
In Sec.(III) we determine the black hole interpretation of the metric, 
and we consider in which sense mass and angular momentum are defined 
and conserved. We shall give a detailed description of the rather 
intricate causal structure and the related Penrose-Carter
diagrams, but we do not 
discuss whether the black holes can result from gravitational 
collapse.\\ 
In Sec.(IV) we describe the rotating toroidal black hole's metric, 
together with an account of its main features, including the causal 
structure and the causal diagrams.\\ 
In this paper we shall use the curvature conventions of Hawking-Ellis'
book \cite{haell} and employ Planck's dimensionless units.
 
\section{Spacetime metric for $g>1$ rotating black holes}

We begin by recalling the uncharged topological black holes discussed in
\cite{mann97,vanz97}. The metric appropriate for genus $g>1$ reads
\begin{eqnarray}
ds^2 = -V(r)dt^2 + V(r)^{-1}dr^2 + r^2(d\theta^2 + \sinh^2\theta d\phi^2)
\label{luciano}
\end{eqnarray}
with the lapse function $V(r)$ given by
\begin{eqnarray}
V(r) = -1 -\frac{\Lambda r^2}{3} - \frac{2\eta}{r}
\end{eqnarray}
where $\eta$ is the mass parameter and $\Lambda=-3\ell^{-2}$ the cosmological
constant. One notices that the $(\theta,\phi)$-sector of the metric describes 
the two-dimensional non-compact space with constant, negative 
curvature. As is well known, this is the universal 
covering space for all Riemannian surfaces with genus $g>1$. Therefore, 
in order to get a compact event horizon, 
suitable identifications in the $(\theta,\phi)$-sector have to be carried 
out, corresponding to the choice of some discrete group of isometries 
acting on hyperbolic $2$-space properly discontinuously. After this 
has been done, the metric 
(\ref{luciano}) will describe higher genus black holes. The $g=1$ 
case, with toroidal event horizon, is given by the metric  
\begin{eqnarray}
ds^2 = -V(r)dt^2 + V(r)^{-1}dr^2 + r^2d\sigma^2
\label{luciano1}
\end{eqnarray}
with the lapse function $V(r)$ given by
\begin{eqnarray}
V(r) =-\frac{\Lambda r^2}{3} - \frac{2\eta}{r}
\end{eqnarray}
and $d\sigma^2$ is the line element of a flat torus. Its conformal structure 
is completely determined by a complex parameter in the upper complex 
half plane, $\tau$, which is known as the Teichm\"{u}ller parameter. A 
rappresentative for the flat torus metric can then be written in the form
\begin{eqnarray}
d\sigma^2=|\tau|^2dx^2+dy^2+2\mathop{\rm Re}\nolimits \tau dx\,dy
\end{eqnarray}  
It is quite trivial to show that all such solutions have indeed a black hole 
interpretation, with various horizons located at roots of the 
algebraic equation $V(r)=0$, provided $\eta$ be larger than some 
critical value depending on $\Lambda$. It can also be shown that for
all genus, 
a ground state can be defined relative to which the ADM mass is 
a positive, concave function \cite{vanz97} of the black hole's temperature as 
defined by its surface gravity \cite{gibb1}, and that the entropy 
obeys the area law\cite{bril97,vanz97}. 

We now determine at least one class of rotating generalizations of the above 
solutions starting with the higher genus case, namely when $g>1$. The 
toroidal rotating black hole will be described last. 
The metric (\ref{luciano}) looks very similar to the Schwarzschild-de Sitter
metric \cite{gibb,carter1,carter2}
\begin{eqnarray}
ds^2 &=& -(1 - \frac{\Lambda r^2}{3} - \frac{2\eta}{r})dt^2
         +(1 - \frac{\Lambda r^2}{3} - \frac{2\eta}{r})^{-1}dr^2 \nonumber \\ 
     & & + r^2(d\theta^2 + \sin^2\theta d\phi^2)  \label{SsdS}
\end{eqnarray}
(Here $\Lambda > 0$).\\ 
For the latter, it is well known that a generalization to the rotating case
exists, namely the Kerr-de Sitter spacetime \cite{gibb,carter1,carter2},
which describes rotating
black holes in an asymptotically de Sitter-space. Its metric reads in
Boyer-Lindquist-type coordinates
\begin{eqnarray}
ds^2 &=& \rho^2(\Delta_r^{-1}dr^2 + \Delta_{\theta}^{-1}d\theta^2) \nonumber \\ 
     & & + \rho^{-2}\Xi^{-2}\Delta_{\theta}[adt - (r^2 + a^2)d\phi]^2\sin^2\theta
         \nonumber \\ 
     & & - \rho^{-2}\Xi^{-2}\Delta_r[dt - a\sin^2\theta d\phi]^2 \label{KdS},
\end{eqnarray}
where
\begin{eqnarray}
\rho^2 &=& r^2 + a^2\cos^2\theta \nonumber \\ 
\Delta_r &=& (r^2 + a^2)(1 - \frac{\Lambda r^2}{3}) - 2\eta r \nonumber \\ 
\Delta_{\theta} &=& 1 + \frac{\Lambda a^2}{3}\cos^2\theta \nonumber \\ 
\Xi &=& 1 + \frac{\Lambda a^2}{3}
\end{eqnarray}
and $a$ is the rotational parameter.\\ 
Now we note that (\ref{luciano}) can be obtained from (\ref{SsdS}) by the
analytical continuation
\begin{eqnarray}
&&t \to it, \; r \to ir, \; \theta \to i\theta, \; \phi \to \phi, \nonumber \\ 
&&\eta \to -i\eta, \label{ancon}
\end{eqnarray}
thereby changing also the sign of $\Lambda$ (this may be interpreted as
an analytical continuation, too).\\ 
Therefore we are led to apply the analytical continuation (\ref{ancon})
also to Kerr-de Sitter spacetime (\ref{KdS}), additionally replacing
$a$ by $ia$. This leads to the metric
\begin{eqnarray}
ds^2 &=& \rho^2(\Delta_r^{-1}dr^2 + \Delta_{\theta}^{-1}d\theta^2) \nonumber \\ 
     & & + \rho^{-2}\Xi^{-2}\Delta_{\theta}[adt - (r^2 + a^2)d\phi]^2\sinh^2\theta
         \nonumber \\ 
     & & - \rho^{-2}\Xi^{-2}\Delta_r[dt + a\sinh^2\theta d\phi]^2, \label{diet1}
\end{eqnarray}
where now
\begin{eqnarray}
\rho^2 &=& r^2 + a^2\cosh^2\theta \nonumber \\ 
\Delta_r &=& (r^2 + a^2)(-1 - \frac{\Lambda r^2}{3}) - 2\eta r \nonumber \\ 
\Delta_{\theta} &=& 1 - \frac{\Lambda a^2}{3}\cosh^2\theta \nonumber \\ 
\Xi &=& 1 - \frac{\Lambda a^2}{3}
\end{eqnarray}
and $\Lambda < 0$.\\ 
One observes that (\ref{diet1}) describes a spacetime which reduces, in 
the limit $a=0$, to the static topological black holes
(\ref{luciano}). 
For our further purpose it is convenient to write (\ref{diet1}) in the 
form
\begin{eqnarray}
ds^2 &=& -\frac{\rho^2\Delta_{\theta}\Delta_r}{\Xi^2\Sigma^2}dt^2+\frac{\rho^2}{\Delta_r}dr^2 + 
          \frac{\rho^2}{\Delta_{\theta}}d\theta^2 \nonumber \\ 
     & & +\frac{\Sigma^2\sinh^2\theta}{\Xi^2\rho^2}[d\phi - \omega dt]^2
         \label{diet2}
\end{eqnarray}
where we introduced
\begin{eqnarray}
\Sigma^2=(r^2+a^2)^2\Delta_{\theta}-a^2\sinh^2\theta\Delta_r
\end{eqnarray}
and the angular velocity
\begin{eqnarray}
\omega=\frac{a[(r^2+a^2)\Delta_{\theta}+\Delta_r]}{\Sigma^2}.
\label{omega}
\end{eqnarray}
Next we show how to compactify the 
$(\theta,\phi)$-sector into a Riemann surface while preserving the 
differentiability of the metric. The timelike $3$-surfaces 
at fixed coordinate radius $r$ are foliated by surfaces at fixed 
coordinate time $t$ into a family of spacelike $2$-surfaces, and
we would like 
these to be Riemann surfaces with genus $g>1$. The metric induced on such 
surfaces is
\begin{eqnarray}
d\sigma^2=\frac{\rho^2}{\Delta_{\theta}}d\theta^2 +
\frac{\Sigma^2\sinh^2\theta}{\Xi^2\rho^2}\,d\phi^2
\label{rieman}
\end{eqnarray}
Note that the Gaussian curvature of this metric is no more constant,
as it was in the case $a=0$.\\
In order to get a Euclidean metric we have to require that
$\Sigma^2>0$. This is the case for every 
$r \in  {\hbox{{\rm I}\kern-.2em\hbox{\rm R}}}$, $\theta \ge 0$, if 
$-a\ell^{-2}(a^2+\ell^2)<\eta <a\ell^{-2}(a^2+\ell^2)$
(or for every $r \ge 0$, $\theta \ge 0$, if $\eta > -a\ell^{-2}(a^2+\ell^2)$).
Outside the prescribed $\eta$ interval, the metric may become singular
or may change the signature.\\
The compactification is now performed in the same way as for a
Riemann surface of constant curvature (i.~e.~for $a = 0$;
in this case see e.~g.~\cite{balasz}). That is, we have to
identify opposite sides of a properly chosen regular geodesic $4g$-gon
centered at the origin $\theta = 0$. The geodesics have to be computed
from the metric (\ref{rieman}), and therefore they are different from those
in the case of constant curvature. The size of the $4g$-gon is determined
by the requirement that the sum of the polygon angles being equal to
$2\pi$ \cite{balasz}, in order to avoid conical singularities.
Indeed, the local version of the
Gauss-Bonnet theorem yields
\begin{equation}
\int_B K\,dA = 2\pi - \sum_{i=1}^{4g}(\pi - \beta_i), \label{gb}
\end{equation}
where $B$ is the interior of the geodesic polygon, $K$ the Gaussian curvature
of the $(\theta,\phi)$-surface, $dA$ the area element of the metric
(\ref{rieman}), and $\beta_i$ the $i$-th polygon angle. (Of course,
the $\beta_i$ are all equal, as the polygon is regular). From (\ref{gb})
we see that the requirement $\sum_i \beta_i = 2\pi$ fixes the size of $B$.
(\ref{gb}) then gives
\begin{equation}
\int_B K\,dA = 2\pi(2 - 2g), \label{gbglob}
\end{equation}
which is the Gauss-Bonnet theorem for a Riemann surface of genus $g$.
A priori, it is not obvious that a polygon which satisfies (\ref{gbglob})
with $g>1$ really exists. Therefore let us sketch a short existence
proof. If the polygon is very small, the sum of the interior angles is
larger than $2\pi$, since the metric (\ref{rieman}) approaches a flat
metric for $\theta \to 0$. On the other hand, enlarging the polygon,
the sum of the angles decreases until it is zero at a certain limit.
(This is the limit when the polygon vertices lie on the border of
the Poincar\'e disc on which (\ref{rieman}) can also be defined by a
proper coordinate transformation. The geodesics meet this border
orthogonally and therefore the angle sum is zero).
As $\sum_i \beta_i$ is a continuous function of the distance of
the vertices from the origin $\theta = 0$,
we deduce that the desired polygon
indeedly exists.\\
The next question which arises is that of the differentiability of the
metric after the compactification. Identifying geodesics assures that
the metric is in $C^1$. (Use "Fermi coordinates" \cite{synge} in
neighbourhoods of the geodesics which have to be identified. In these
coordinates one has on the geodesics
$\sigma_{ij} = \delta_{ij}$ and $\sigma_{ij,k} = 0$). As the second
derivatives of the metric are bounded, the metric is even in $C^{1,\alpha}$
(which means that the first derivatives are Hoelder continuous with exponent
$\alpha$). 
Now we note that one obtains the
Gaussian curvature $K$ by applying a quasilinear elliptic operator $L$
of second order to the metric $\sigma$,
\begin{equation}
K = L[\sigma].
\end{equation}
$L$ can be written as
\begin{equation}
L = \sum_{\beta \le 2} a_{\beta}(x,\partial^l \sigma)\partial^{\beta},
\end{equation}
where $x$ stands for the coordinates on the surface, $\beta$ is a
multi--index, the coefficients $a_{\beta}$ are matrices, and $l \le 1$.
We now express the zeroeth and first derivatives of the metric in
$a_{\beta}(x,\partial^l \sigma)$ as functions of the coordinates $x$.
This makes the operator $L$ linear with coefficients in $C^{0,\alpha}$.
As the Gaussian curvature is also in
$C^{0,\alpha}$ on the compactified surface, we conclude from the
regularity theorem for solutions of linear elliptic equations
\cite{besse,gilbarg} that the metric $\sigma$ is (at least) in
$C^{2,\alpha}$.\\
As we have compactified now the $(\theta,\phi)$-sector to a Riemann
surface $S_g$, the topology of the manifold
is that of ${\hbox{{\rm I}\kern-.2em\hbox{\rm R}}}^2\times S_g$.\\
Finally, we remark that (\ref{diet1}) is a limit case of the metric of
Plebanski and Demianski \cite{pleb}, which is the most general known 
Petrov type-D solution of the source-free Einstein-Maxwell equations 
with cosmological constant. In the case of zero electric and magnetic 
charge it reads
\begin{eqnarray}
ds^2 &=& \frac{1}{(1 - pq)^2}\left\{\frac{p^2 + q^2}{\cal P}dp^2 +
         \frac{\cal P}{p^2 + q^2}(d\tau + q^2 d\sigma)^2\right. \nonumber \\ 
     & & \left.+ \frac{p^2 + q^2}{\cal L}dq^2 -
         \frac{\cal L}{p^2 + q^2}(d\tau - p^2 d\sigma)^2 \right\}, \label{PD}
\end{eqnarray}
where the structure functions are given by
\begin{eqnarray}
\cal P &=& \left(-\frac{\Lambda}{6} + \gamma\right) + 2np - \epsilon p^2
           + 2\eta p^3 + \left(-\frac{\Lambda}{6} - \gamma\right)p^4
           \nonumber \\ 
\cal L &=& \left(-\frac{\Lambda}{6} + \gamma\right) - 2\eta q + \epsilon q^2
           - 2n q^3 + \left(-\frac{\Lambda}{6} - \gamma\right)q^4.
\end{eqnarray}
$\Lambda$ is the cosmological constant, $\eta$ and $n$ are the mass and nut
parameters, respectively, and $\epsilon$ and $\gamma$ are further
real parameters. (For details c.f. \cite{pleb}). Rescaling the coordinates
and the constants according to
\begin{eqnarray}
&&p \to L^{-1}p, \; q \to L^{-1}q, \; \tau \to L\tau, \; \sigma \to L^3\sigma,
  \nonumber \\ 
&&\eta \to L^{-3}\eta, \; \epsilon \to L^{-2}\epsilon, \; n \to L^{-3}n, \;
  \gamma \to L^{-4}\gamma + \frac{\Lambda}{6}, \; \Lambda \to \Lambda
\end{eqnarray}
and taking the limit as $L \to \infty$, one obtains
\begin{eqnarray}
ds^2 &=& \frac{p^2 + q^2}{\cal P}dp^2 +
         \frac{\cal P}{p^2 + q^2}(d\tau + q^2 d\sigma)^2 \nonumber \\ 
     & & + \frac{p^2 + q^2}{\cal L}dq^2 - \frac{\cal L}{p^2 + q^2}(d\tau -
         p^2 d\sigma)^2,
\end{eqnarray}
where now
\begin{eqnarray}
\cal P &=& \gamma + 2np - \epsilon p^2 - \frac{\Lambda}{3}p^4 \nonumber \\ 
\cal L &=& \gamma - 2\eta q + \epsilon q^2 - \frac{\Lambda}{3}q^4.
\end{eqnarray}
Setting now
\begin{eqnarray}
&&q = r, \; p = a\cosh\theta, \; \tau = \frac{t - a\phi}{\Xi}, \; \sigma = 
  -\frac{\phi}{a\Xi}, \nonumber \\ 
&& \epsilon = -1 - \frac{\Lambda a^2}{3}, \; \gamma = -a^2, \; n = 0,
\end{eqnarray}
one gets our solution (\ref{diet1}).
As we said, the metric (\ref{diet1}) results to be a
limit case of the more general
solution (\ref{PD}) of Einstein's equation.
This formally shows that (\ref{diet1}) should solve Einstein's 
field equations with cosmological constant, i.e. the analytical
continuation (\ref{ancon}) of the Kerr-de Sitter metric should yield again
a solution. Anyhow, one could doubt of the procedure as it involves an
infinite limit of some parameters in the initial solution of Einstein's 
equations. Therefore, let us sketch a short independent 
proof of the fact that (\ref{diet1}) still satisfies  Einstein's 
equations.\\ 
Generally speaking, all functions which appear in the 
left hand side of  Einstein's equations containing the cosmological
constant are polynomial in  metric tensor components, components of
the inverse metric tensor and derivatives of metric tensor components.
Considering all these functions  as independent variables, the l.h.s of
Einstein's equations defines analytic functions in these variables.
Let us consider Kerr-de Sitter spacetime defined above. Then the metric,
its inverse and its derivatives define locally analytic functions 
of the (generally complex) variables $t, r, \theta, \phi, \eta, a, \Lambda$.
We conclude that the
l.h.s of Einstein's equations defines 
analytic functions of  $(t, r, \theta, \phi, \eta, a, \Lambda)$
in open connected domains away from singularities corresponding to 
zeroes of $\Delta_r$ and the determinant of the analitically continued 
metric ($g= -\Xi^{-4} (r^2 + a^2 \cos^2\theta)^2 $).
Moreover, we know that, for real values of $(t, r, \theta, \phi, \eta, 
a, \Lambda)$, $\Lambda >0$, these functions  vanish because 
Kerr-de Sitter metric is a solution of Einstein's  equations. 
Hence, due to the theorem of uniqueness of the analytical 
continuation of a function of several complex variables, they must 
vanish  concerning all (generally complex) remaining values
of  $(t, r, \theta, \phi, \eta, a, \Lambda)$, provided they belong to the same
domain of analyticity of the previously considered real values.
In particular, we can pick out the set of values determining the metric 
(\ref{diet1}) as final values. 
Notice that these values belong to the same analyticity domain of
the values determining 
Kerr-de Sitter metric because one can easily find piecewise 
smooth trajectories in the space 
$ {\ \hbox{{\rm I}\kern-.6em\hbox{\bf C}}}^7$, connecting Kerr-de Sitter parameters 
to parameters appearing in the 
metric (\ref{diet1}), and skipping all singularities.

\section{Some properties of $g>1$ rotating black holes}

We shall briefly discuss now the black hole interpretation of the 
proposed solutions and some of their physical 
properties, starting with the case $g>1$. 

\subsection{Curvature}
Let us begin by looking at the curvature of the spacetime
metric (\ref{diet1}). 
The only nonvanishing complex tetrad component of the Weyl tensor is
given by
\begin{eqnarray}
\Psi_2 = -\frac{2\eta}{(r + ia\cosh\theta)^2}
\end{eqnarray}
(The $\Psi_i$, $i = 0,\ldots, 4$, are the standard complex tetrad
components describing the conformal curvature. For details c.f.
\cite{pleb}, \cite{kram}). 
For $\eta = 0$ the Weyl tensor vanishes
and, since $R_{ij} = \Lambda g_{ij}$, our manifold is a space of constant
curvature, $k=-\ell^{-2}$, i.e. a quotient space of the universal covering of
anti-de Sitter space. This situation is comparable to that of the Kerr metric,
which, for vanishing mass parameter, is simply the Minkowsky metric written
in oblate spheroidal coordinates.\\ 
One further observes that $\Psi_2$ is always
nonsingular, in particular the curvature singularity in Kerr-de Sitter
space at $\rho^2 = 0$, i.e. $r=0, \theta = \pi/2$ vanishes after the
analytical continuation, as $r^2 + a^2\cosh^2 \theta$ is always positive.
Hence the manifold may be extended to values $r < 0$, and closed 
timelike curves will always be present. This is similar
to the BTZ black hole \cite{btz}, where no curvature singularity occurs 
(see also \cite{amin1} for an exhaustive determination of $(2+1)$-black 
holes and their topology). On the other hand all non rotating
solutions with $\eta\neq0$ found so far have curvature singularities at 
the origin, but do not violate the strong causality condition.

\subsection{Singularity structure and horizons}

The metric (\ref{diet1})
becomes singular at $\Delta_r = 0$. With $\Lambda = -3\ell^{-2}$ this equation reads
\begin{eqnarray}
(r^2 + a^2)\left(\frac{r^2}{\ell^2} - 1\right) - 2\eta r = 0 \label{zeroes}
\end{eqnarray}
There are several cases, in all of which $\Delta_r$ is
positive for $r$ smaller than the left most zero or larger
than the right most zero. 
Here the various cases:\\ 
$i)$ If $\ell^2 < a^2(7 + 4\sqrt{3})$ and $\eta\in {\hbox{{\rm I}\kern-.2em\hbox{\rm R}}}$
there is only one positive
solution $r_+$ of (\ref{zeroes}) and only one negative solution $r_-$. For
$r > r_+$ and $r < r_-$, $\Delta_r$ is positive and
$\partial_r$ is spacelike. For $r_- < r < r_+$, $\partial_r$ becomes
timelike. $r_-$ and $r_+$ are first order zeroes. The causal structure 
on the axis is given in Fig.~(1).\\ 
$ii)$  If $\ell^2 = a^2(7 + 4\sqrt{3})$ and\\ 
$a)$  $\eta \neq \pm\eta_0  $, $\eta_0= (4\ell/3)(26\sqrt{3}-45)^{1/2}  >0$,
then the solutions behave as in the case
$(i)$;\\ 
$b)$  $ \eta = -\eta_0$, then there is a first order
root $r_{-}$
for $r<0$ and a third order root $r_+ = [(\ell^2-a^2)/6]^{1/2}$
for $r>0$ and $\Delta_r$ changes sign by crossing the roots;\\
$c)$  $ \eta = \eta_0$, then there is a first order
root $r_{+}$
for $r>0$ and a third order root $r_- = -[(\ell^2-a^2)/6]^{1/2}$
for  $r<0$. Again $\Delta_r$ changes sign crossing the roots.
The Penrose-Carter diagrams in the cases $b)$ and $c)$ are also given by
Fig.~(1).\\ 
$iii)$ If $\ell^2 > a^2(7 + 4\sqrt{3})$ and $\eta\leq 0$
we have again several subcases. Let
\begin{eqnarray}
R_{\pm} &=& \sqrt{\frac{1}{6}(\ell^2 - a^2) \mp \frac{1}{6}
\sqrt{(\ell^2 - a^2)^2
             - 12\ell^2a^2}} \nonumber \\ 
\eta_{\pm} &=& -\frac{2R_{\pm}}{3\ell^2}\left[(\ell^2 - a^2) \pm \frac{1}{2}
               \sqrt{(\ell^2 - a^2)^2 - 12\ell^2a^2}\right]
\end{eqnarray}
(Note that $\eta_- < \eta_+ < 0$).\\ 
$a)$ for $0\geq \eta > \eta_+$ $\Delta_r$ behaves as in $(i)$;\\ 
$b)$ for $\eta = \eta_+$, $\Delta_r$
has two positive zeroes $r_+=R_+$ and $r_{++} > r_+$ and a negative
zero $r_{--}$.  At
$r=r_+$ the graph of $\Delta_r$ versus $r$ is tangent to the
$r$-axis and $\Delta_r$ does not change sign ($r_+$ is a second order zero),
whereas at $r=r_{++}$ and
$r=r_{--}$,
$\Delta_r$ changes sign from negative to positive values 
and from positive to negative values respectively, as $r$ increases. These 
are first order zeroes.
The causal structure is shown in Fig.~(3);\\ 
$c)$ for $\eta_- < \eta < \eta_+$, $\Delta_r$ has three positive zeroes
$r_-$, $r_+$, $r_{++}$ and one
negative zero $r_{--}$ where $\Delta_r$ changes sign.
These zeroes are first order; 
the causal structure is shown in Fig.~(2).\\ 
$d)$ in the case $\eta = \eta_-$
one obtains again two positive
roots $r_-$ and $r_{++}=R_-> r_-$, and a negative root $r_{--}$.
At $r_-$ and $r_{--}$, which are first order roots,
$\Delta_r$ changes
sign from $-$ to $+$ and from $+$ to $-$ respectively,
whereas at $r_{++}$, which is a second order root,
$\Delta_r$ does not change sign.
For the corresponding Penrose-Carter diagram see Fig.~(4);\\ 
$e)$ for $\eta < \eta_-$ we get again the same behaviour as in
$(i)$.\\ 
$iv)$ If $\ell^2 > a^2(7 + 4\sqrt{3})$ and $\eta > 0$ the discussion of the
roots is symmetric to that for the case $(iii)$, 
considering the symmetry of (\ref{zeroes})
under the combined inversion $r\rightarrow -r$, $\eta \rightarrow -\eta$.
In this case one has in general one positive first order zero and up to
three negative zeroes.\\
All zeroes of $\Delta_r$ in the examined cases are merely coordinate
singularities, similar to
the Schwarzschild case. They represent horizons, as the normals to
the constant $t$ and constant $r$ surfaces become null when $r$ is a
root of $\Delta_r=0$. The pair of outermost horizons $r_H$
(e.~g.~$r_H=r_{++}$ and
$r_H= r_{--}$ in case $(iii,d)$) are also
event horizons as the Killing trajectories in the exterior
stationary domains never
intersect the surfaces $r=r_H$. The future parts of these
event horizons are the boundary of
the causal past of all time-like inextendible geodesics contained in the 
respective stationary regions which reach the future time-like infinity 
(see the Penrose-Carter diagrams). \\ 
However, the resulting causal structure is
rather intricate. We notice the complete absence of metric singularities
at $r=0$. This allows one to consider the coordinate $r$ in the complete
range $(-\infty,+\infty)$ as we did above.\\ 
We remark that there is an extreme case ((iii,d)) which for $a \to 0$
gives the naked singularity discussed in \cite{mann97,vanz97}, but
for $a>0$ still represents a black hole. Hence the
non rotating naked singularity is unstable, as it turns into a black hole
by an infinitesimal addition of angular momentum. This seems to lend
some support to the cosmic censorship conjecture.

In all cases discussed above, the outermost zeroes represent
event horizons. Their Gaussian curvature is given by
\begin{eqnarray}
K = -\frac{1}{\rho_H^6}\left[(\rho^2_H - 4a^2\cosh^2\theta)\left(
         (r_H^2 + a^2)\Delta_{\theta} + \frac{\rho^2_H a^2}{l^2}
         \sinh^2\theta\right)
         + \frac{4a^2 \rho^4_H}{l^2}\cosh^2\theta \right],
\end{eqnarray}
where the index $H$ indicates that the corresponding quantities are to be
evaluated on the event horizon $r_H$. $K$ is no more constant as in the
nonrotating case, because the
horizon has been warped by the rotation.

\subsection{Angular velocity and surface gravity}

At least for $\eta > - a \ell^{-2}(a^2 + \ell^2)$,
the positive event horizon (as well as any $r= constant > 0$ surface)
rotates relative to 
the stationary frame at infinity, where $\partial_t$ is time-like,
 with angular velocity 
$\Omega_H=\omega(r_H,\theta)$, where $\omega$ is given by Eq.~(\ref{omega}), which yields 
\begin{eqnarray}
\Omega_H=\frac{a}{r_H^2+a^2}
\end{eqnarray}
Notice that $\omega(r,\theta)$ is just 
given by $d\phi/dt$ along time-like trajectories with fixed values 
for $r$ and $\theta$, $t$ being proportional to the proper time 
$\tau$ according to $t = (\Xi \Sigma/\rho \sqrt{\Delta_\theta\Delta_r})
\tau $. These are trajectories of co-rotating observers.\\ 
There also exists a dragging effect at infinity, as $\omega$ is non 
vanishing there, its value being $\Omega_{\infty}=a/(a^2+\ell^2)$.\\ 
The surface gravity $\kappa$ is another important
property of the event horizon. 
It is normally defined in terms of the null, future pointing 
generators of the horizon, using
\begin{eqnarray}
l^c\nabla_cl^a=\kappa l^a \qquad \l^a=\partial_t+\Omega_H\partial_{\phi}
\end{eqnarray}
However, although in the present case $\partial_t$ still is a
global Killing field, 
the vector $\partial_{\phi}$ is only a local Killing field, because 
of the procedure used to build up $S_g$.
This agrees with the known result that Riemann 
surfaces with $g>1$ admit no global Killing fields, nor even 
global conformal Killing 
fields. Nevertheless, the surface gravity can still be defined as the 
acceleration per unit coordinate time which is necessary to hold in 
place a co-rotating particle (i.e. one at some fixed $r$ and $\theta$)
near the 
event horizon. Such a particle will move on the trajectories 
considered above, where $\omega = d\phi/d t$. These trajectories
are integral curves of the vector field 
$u=N^{-1}(\partial_t+\omega\partial_{\phi})$, which is timelike 
everywhere in the $r>0$  
exterior domain  bounded by the outermost event horizon.
Notice that $Nu$ is a timelike Killing field and thus the exterior domain
is stationary.                                                      
The 
function $N$ normalizing the four-velocity is the lapse function of 
the foliation determined by the Killing coordinate time $t$, and is
\begin{eqnarray}
N^2=\frac{\rho^2\Delta_{\theta}\Delta_r}{\Xi^2\Sigma^2}
\end{eqnarray}
By computing the four-acceleration, one obtains in this way
\begin{eqnarray}
\kappa=\frac{1}{2(a^2+\ell^2)(r_H^2+a^2)}\left[3r_H^3+(a^2-\ell^2)r_H
+\frac{a^2\ell^2}{r_H}\right]
\end{eqnarray}
Remarkably, this is constant over the event horizon even in the absence 
of a true rotational symmetry. In view of this last fact, the meaning of 
the surface gravity as the quantum temperature of the black hole remains 
a little bit obscure. The fact is that, although 
one can define a conserved mass by using the time translation symmetry 
of the metric, one cannot define a strictly conserved angular 
momentum, but only a conserved angular momentum with respect to a 
special choice of the observers at infinity. Hence the status of the 
first law for such black holes certainly needs further clarifications. 
As we will see, the situation will be rather different for toroidal 
black holes, which behave quite similarly to the Kerr solution. This 
also suggests that higher genus rotating black holes may be a kind of 
stable soliton solutions in anti-de Sitter gravity.\\ 
>From the metric (\ref{diet1}) we may read off
\begin{eqnarray}
g_{tt} = \frac{a^2 \Delta_{\theta} \sinh^2 \theta -\Delta_{r}}{\rho^2 \Xi^2}
\end{eqnarray}
>From this expression, one recognizes that
$g_{tt}$ may change sign within all regions where
$\Delta_r > 0$.
For that reason one cannot define "static" co-moving observers
with the coordinates $t,r,\theta,\phi$ near the outermost horizons 
but only "non-static" co-rotating observers as we did above.
Anyhow, $g_{tt} < 0$ for $|r|$ sufficiently large.
The surface where $g_{tt}=0$ inside any region where $\Delta_r > 0$ 
is one of the boundaries of an ergoregion in which
both $\partial_t$ and $\partial_r$ are spacelike. 
This is therefore a stationary limit surface, locally determined by
\begin{eqnarray}
a^2 \Delta_{\theta} \sinh^2{\theta} = \Delta_r.
\end{eqnarray}
The remaining boundaries of this ergoregion are
event horizons located at roots of $\Delta_r$. 
These are general features of
rotating black hole metrics. Furthermore, similarly
to the Kerr solution, the event horizon and the surrounding stationary 
limit surface meet at $\theta=0$, where they are smoothly tangent to each 
other provided $\Delta_r$ vanishes in a first-order zero. 

\subsection{Mass and angular momentum}

The two conserved charges which are associated with a 
rotating self-gravitating system are the mass and the 
angular momentum.\\ 
One approach to a general and sensible definition of conserved charges 
associated to a given spacetime, is the canonical ADM analysis 
appropriately extended to include non-asymptotically flat solutions. 
This led to the introduction of the more general concept of 
quasi-local energy \cite{york93} for a spatially bounded 
self-gravitating system, and more generally, to various other
quasi-local conserved charges. These may be obtained as follows. One 
considers a spacetime enclosed into a timelike three-boundary $B$, 
which is assumed to be orthogonal to a family of spatial slices, 
$\Sigma_t$, foliating spacetime. The slices foliate the boundary into a 
family of $2$-surfaces $ {\cal B}_t=\Sigma_t\cap B$ (which need not be 
connected), and these will have outward 
pointing spacelike normals in $\Sigma_t$, denoted $\xi^a$, and future 
pointing normals in $B$, denoted $u^a$. 
The validity of vacuum Einstein's equations in the inner 
region, with or without cosmological constant, 
then implies along $B$ the usual diffeomorphism constraint of general 
relativity
\begin{eqnarray}
D_a(\Theta^{ab}-b^{ab}\Theta)=0  \qquad \Theta=\Theta^{a}_{\,a}
\label{claw}
\end{eqnarray}
where $b_{ab}$ is the boundary three-metric, $D_a$ the 
associated covariant derivative along $B$ and $\Theta_{ab}$ its extrinsic 
curvature. If now the boundary three-metric admits a Killing vector 
$K^a$, then contracting (\ref{claw}) with $K_a$ and integrating over 
$B$ from $ {\cal B}_{t_1}$ to $ {\cal B}_{t_2}$ one obtains the conservation 
law $Q_{K}(t_1)=Q_{K}(t_2)$, where the conserved charge is
\begin{eqnarray}
Q_K(t)=-\frac{1}{8\pi}\int_{ {\cal B}_t}[\Theta_{ab}-b_{ab}\Theta]
K^au^b\sqrt{\sigma}
\label{charge}
\end{eqnarray}
The quasi-local mass is then defined to be the charge associated with 
the time evolution vector field of the foliation $\Sigma_t$, when this is 
a symmetry of the boundary geometry. This field will be 
$K^a=Nu^a+V^a$, with lapse function $N$ and the shift vector $V^a$ 
constrained to be tangent to $B$. In this way the time evolution of 
the three-geometry on $\Sigma_t$ induces a well defined time evolution of the 
two-geometry of $ {\cal B}_t$ along $B$. The quasi-local energy is defined 
for observers which travel orthogonally to $ {\cal B}_t$ in $B$, i.e. for 
$K^a=u^a$ and is, from Eq.~(\ref{charge})
\begin{eqnarray}
E( {\cal B}_t)=-\frac{1}{8\pi}\int_{ {\cal B}_t}[\Theta_{ab}u^au^b+\Theta]\sqrt{\sigma}
\label{qulo}
\end{eqnarray}
It can be shown that $E({\cal B}_t)$ is minus the rate of change of the 
on-shell gravitational action per unit of {\it proper time} along the timelike 
boundary $B$ \cite{york93}, a fact which motivates the definition. 
However as $u^a$ is not in general a 
symmetry of the boundary, the quasi-local energy in not conserved, e.g. 
gravitational waves may escape from the region of interest, and it can 
also be negative (binding energy, c.f. \cite{hayw}). In our 
non asymptotically flat context, where the lapse function diverges at 
infinity, one can define the quasi-local energy by measuring 
the rate of change of the action per unit of {\it coordinate time}. 
Then one uses $K^a=Nu^a$ and the energy is as in Eq.~(\ref{qulo}) but with 
a further factor $N$ under the integral, so we denote it by 
$E_N( {\cal B}_t)$. 
Similarly, the angular momentum will be the charge associated to a rotational 
symmetry, generated by a spacelike Killing field $\tilde{K}^a$.\\ 
It is very important that York and Brown's quasi-local charges be functions 
of the canonical data alone. If background subtractions were 
necessary, these ought to be chosen appropriately to achieve this 
requirement. The quasi-local mass can also be arrived at 
by a careful handling of the 
boundary terms in the Hamiltonian for general relativity. Then one arrives 
at the equivalent expression for the mass \cite{york93,regge,hawk1}, 
as measured from infinity
\begin{eqnarray}
M=-\frac{1}{8\pi}\int_{S_g(R)}[N(\tilde{\Theta}-\tilde{\Theta}_0)-16\pi(P_{ab}V^a\xi^b
-(P_{ab}V^a\xi^b)_{|0})]\sqrt{\sigma}\,d^2x
\label{mass}
\end{eqnarray}
where quantities with a subscript $0$ denote background subtractions, 
chosen so that $M$ be a function of the canonical data 
alone\cite{york93}, and the limit $R\to\infty$ is understood.\\ 
In our case, $S_g(R)$ is an asymptotic Riemann surface at $r=R$ 
embedded in a $t=constant$ slice, with outward pointing normal $\xi^a$
and extrinsic curvature $\tilde{\Theta}$, $P_{ab}$ is the momentum 
canonically conjugate to the metric induced on the slice, and $(N,V^a)$ are
the lapse function and the shift vector of the $t=constant$ 
foliation.\\
The charge associated to a rotational Killing symmetry generated by $\tilde{K}^a$, 
can also be written as a function of the canonical 
data, and is
\begin{eqnarray}
J=-2\int_{S_g}[P_{ab}\tilde{K}^a\xi^b-(P_{ab}\tilde{K}^a\xi^b)_{|0}]
\sqrt{\sigma}\,d^2x
\label{momen}
\end{eqnarray}
Unlike the case of non-rotating topological black holes, where a 
natural choice for the background can be made, no distinctive 
background metric has been found in the present case. The best we are 
able to do is to define the mass relative to some other solution with 
the same topology and rotation parameter. In spite of the dragging 
effect at infinity and the intricate form of the metric, 
what we get is the very simple result
\begin{eqnarray}
M=\frac{\eta - \eta_0}{4\pi\Xi(r^2_H+a^2)}{\cal A}_H,
\end{eqnarray}
where ${\cal A}_H$ is the horizon area.
$\eta$ can be expressed as a function of the outermost horizon 
location $r_H$, by using $\Delta_r(r_H,\eta)=0$. Thus $\eta$ really is related 
to the Hamiltonian mass, albeit in a relative sense. The 
quasi-local energy is not equal to the quasi-local mass and is not 
even equal to the mass in the limit $R\to\infty$, a consequence of 
the dragging effect at infinity. Indeed, we obtain the trivial 
result that $E(R)-E_0(R)=0$, if the background has the same rotation 
parameter but different $\eta$. Thus all solutions with equal $a$ have 
the same quasi-local energy.\\ 
Concerning the
angular momentum we are in a different position, since there is no 
global rotational 
Killing symmetry. However the vector $\tilde{K}=\partial_{\phi}$, although it 
is not a Killing vector, obeys locally
the condition $\nabla_{(a}\tilde{K}_{b)}=0$ and 
is therefore a kind of approximate symmetry, we could say 
a locally exact symmetry. We 
may try to compute $J$ using Eq.~(\ref{momen}) with
$\tilde{K}=\partial_{\phi}$. 
Then one finds that $J$ is already finite without any subtraction 
and we get $J=\Xi^{-2}\eta a\, {\cal I}$, where the integral
\begin{eqnarray}
{\cal I}=\frac{3}{8\pi}\int_{S_g}\sinh^3\theta\,d\theta\,d\phi
\end{eqnarray}
has to be performed over a fundamental domain of the Riemann surface 
$S_g$ (we were unable to do this, however). 
This is weakly conserved in the sense that it depends on the 
choice of a spatial slice in the three-boundary at infinity. Due to 
these facts, the 
"first law" and the full subject of black hole thermodynamics needs here 
further clarifications. In this connection, one should use a kind of 
quasi-local formalism for black hole thermodynamics, along the lines 
of Brown et al.~\cite{brow} for asymptotically anti-de Sitter black 
holes.\\ 
One may note, among other things, that $J=0$ for the locally anti-de 
Sitter solution corresponding to $\eta=0$, in agreement with Holst's and
Peldan's theorem\cite{holst97}. Physically, in this case 
$\Omega_H=\Omega_{\infty}$ and the 
horizon does not rotate relative to the stationary observers at infinity. 

\section{The rotating toroidal black hole}

We discuss now another black hole solution in anti-de Sitter gravity 
which represents a rotating torus hidden by an event horizon. The 
first solution of this kind has been discovered by Lemos and Zanchin 
\cite{lemzan} by compactifying a charged open black string. This is a 
solution that can be obtained from the non rotating toroidal metric by 
mixing time-angle variables into new ones. This is not a permissible 
coordinate transformation in the large, as angles, unlike time, are 
periodic variables. This is why the solutions one obtains are 
globally different, 
as clearly showed by Stachel while investigating the gravitational 
analogue of the Aharanov-Bohm effect \cite{stac}.\\ 
The metric we shall present cannot be obtained by forbidden coordinate 
mixing, but it can be obtained from the general Petrov type-$D$ solution 
already presented by a simple choice of parameters. 
By requiring the existence of the non-rotating solution (which we know 
to exist) and the time inversion symmetry, $t\to-t$, $\phi\to-\phi$, 
we get the following metric tensor
\begin{eqnarray}
ds^2=-N^2dt^2+\frac{\rho^2}{\Delta_r}\,dr^2+\frac{\rho^2}{\Delta_P}
\,dP^2+\frac{\Sigma^2}{\rho^2}\,(d\phi-\omega\,dt)^2
\label{torobh}
\end{eqnarray}
where $P$ is a periodic variable with some period $T$, $\phi$ is 
another angular variable with period $2\pi$ and
\begin{eqnarray}
\rho^2=r^2+a^2P^2 \qquad \Delta_P=1+\frac{a^2}{\ell^2}\,P^4
\end{eqnarray}
\begin{eqnarray}
\Delta_r=a^2-2mr+\ell^{-2}r^4 \qquad \Sigma^2=r^4\Delta_P-a^2P^4\Delta_r
\end{eqnarray}
Finally, the angular velocity and the lapse are given by, respectively
\begin{eqnarray}
\omega=\frac{\Delta_rP^2+r^2\Delta_P}{\Sigma^2}\,a \, , \qquad 
N^2=\frac{\rho^2\Delta_P\Delta_r}{\Sigma^2}.
\end{eqnarray}
The solution is obtained as a limit case of the Plebanski-Demianski
metric by setting 
$\varepsilon=0$, $\gamma=a^2$ and rescaling $p=aP$ (this last to
have the limit $a\to 0$).\\ 
The metric induced on the spacelike two-surfaces at some constant $r$ and $t$ 
is then
\begin{eqnarray}
d\sigma^2=\frac{a^2P^2+r^2}{\Delta_P}\,dP^2+
\frac{\Sigma^2}{a^2P^2+r^2}\,d\phi^2
\end{eqnarray}
As long as $\Sigma^2>0$, 
this is a well defined metric on a cylinder, but as it 
stands it cannot be defined on the torus which one gets identifying some 
value of $P$, say $P=T/2$, with $P=-T/2$. This is because the components of 
the metric are even, rational functions of $P$ but have unequal derivatives at 
$\pm T/2$.
Thus we need to cover $S^1\times S^1$ with four coordinate 
patches, and set $P=\lambda\sin\theta$ in a neighborhood of $\theta=0$ and 
$\theta=\pi$ and $P=\lambda\cos\theta$ in a neighborhood of $\theta=\pi/2$ and 
$\theta=3\pi/2$, where $\lambda$ is a constant needed to match the length of 
the circle to the chosen value $T$. 
On the overlap $\cos\theta$ is a $C^{\infty}$ function of 
$\sin\theta$ and viceversa, so now the metric is well defined and smooth on a 
torus.\\ 
Even on the cylinder, the metric (\ref{torobh}) represents a rotating 
cylindrical black hole not isometric to the one discussed by Lemos 
\cite{lemzan,lemocy} or Santos \cite{santo},
which are stationary generalizations 
of the general static cylindrical solution found by Linet 
\cite{linet}. Thus in this case we have not a unique solution, 
but rather a many-parameter family of stationary, locally static 
metrics. This was to be expected as whenever the first Betti number 
of a static manifold is non vanishing, there exists in general a 
many-parameter family 
of locally static, stationary solutions of Einstein's equations,
a fact which can be regarded as a gravitational analogue 
of the Aharanov-Bohm effect \cite{stac}.\\ 
We shall study now the metric (\ref{torobh}) for $m>0$. Notice the symmetry
under the combined inversion $r\rightarrow -r$, $m\rightarrow -m$.
The metric coefficients are functions of $(r,P)$ and $P$ is 
identified independently of $\phi$. Therefore the metric has a 
global rotational symmetry (unlike the higher genus solutions) 
and is stationary. We shall consider mostly 
the region $r \geq 0$ which has a black hole interpretation
and is the physically relevant region for black holes forming
by collapse. 
Anyhow, the metric (\ref{torobh}) admits a sensible continuation to 
$r<0$.

\subsection{Singularity and horizons}

The event horizons arise from the zeroes of $\Delta_r$. In the case
$m>0$ that we are considering, all zeroes may appear in the region
$r\geq 0$ only (see Fig.~(1) for the causal structure in the non 
extreme case). 
Considering the metric (\ref{torobh}), one finds that there is a critical
value, $a_c$, for the rotation parameter $a$, such that for $a>a_c$ the
solution is a naked singularity. For $0\leq a<a_c$ there are two
positive first order roots,
$r_+$ and $r_-$ with $r_+\geq r_-$, which coalesce at
the second order root
$r_+=r_-=(m\ell^2/2)^{1/3}$ when $a=a_c$. This critical value
is
\begin{eqnarray}
a_c=\sqrt{3}(m/2)^{2/3}\ell^{1/3}
\end{eqnarray}
The event horizon is located at the larger value $r_+$, and has a 
surface gravity
\begin{eqnarray}
\kappa=\frac{2r_+^3-m\ell^2}{\ell^2r_+^2}
\end{eqnarray}
The surface gravity vanishes when $a=a_c$ and the metric describes an 
extreme black hole (see Fig.~(5)). 
Finally, there is a curvature singularity at $\rho^2=0$, 
namely at $r=P=0$. As a point set at fixed time, this is $\{p,q\}\times S^1$, 
where $\{p,q\}$ are the two points on the torus at $r=0$ which correspond to 
$P=0$, and looks like a pair of disjoint ring singularities. 
Another point of interest is that 
$g_{tt}>0$ and $\Sigma^2<0$ in a neighborhood of $r=0$. Therefore the 
torus turns into a Lorentzian submanifold with $\phi$ becoming a timelike 
coordinate. Evidently there are closed timelike curves around the 
origin. As we can see, the situation is quite similar to 
the Kerr metric, except that the Euler characteristic of the horizon 
now vanishes. To check this, notice that the metric on the horizon is, 
locally
\begin{eqnarray}
d\sigma^2=\frac{a^2P^2+r_+^2}{\Delta_P}\,dP^2+
\frac{r_+^4\Delta_P}{a^2P^2+r_+^2}\,d\phi^2
\end{eqnarray}
This metric can be written in conformally flat 
form by factoring out the $\phi\phi$-component, which is smooth and 
positive. The conformal metric has $\sigma_{\phi\phi}=1$ and it turns out 
to be flat. The actual metric is thus
conformally flat and defined on a compact domain. The scalar curvature 
of a conformally flat manifold is a total divergence and 
vanishes when integrated over a closed manifold.  
Therefore the Euler characteristic vanishes and the horizon, 
which we assumed to be compact and orientable, must be a torus. 
Furthermore, by rescaling the metric with a constant parameter $\mu$, 
we can see that the periods scale as $2\pi\to 2\pi\mu$, $T\to\mu 
T$. Therefore it is the ratio of the periods that is conformally invariant. 
This ratio determines the conformal class of the torus and is the 
analogue of the more familiar Teichm\"{u}ller parameter.\\ 
Since all surfaces at constant $r$ take on the topology of a torus, 
${\cal T}^2$, the topology of the external region (the domain of outer 
communication in Carter's language) is that of $ {\hbox{{\rm I}\kern-.2em\hbox{\rm R}}}^2 \times {\cal T}^2$.\\ 
Finally, few comments on the presence of ergoregions are in order.
Consideration of the metric (\ref{torobh}) lead us to
\begin{eqnarray}
g_{tt} = \frac{a^2\Delta_P- \Delta_r}{\rho^2} \label{gtttoro}.
\end{eqnarray}
>From this expression, we see that $g_{tt} > 0$ within the
regions where $\Delta_r <0$. Conversely, within regions where
$\Delta_r >0$, outside the external horizon
in particular, $g_{tt}$
may change its sign becoming positive for $|r|$ sufficiently small and
negative for large $|r|$.
In fact, as in the previously considered case, ergotori appear
within the regions  $\Delta_r>0$. In particular, this happens
outside the outermost event horizon. Ergoregions, where both
$\partial_t$ and $\partial_r$ are space-like,
are bounded by event horizons and the surfaces
at $g_{tt}=0$, given by the implicit equation 
\begin{eqnarray}
P^4 = \frac{r^4-2m\ell^2r}{a^4}
\label{ergotoro}.
\end{eqnarray}
Differently from the previously examined class of topological rotating
black holes, surfaces at $g_{tt}=0$ and
horizons do not meet in the  case of a toroidal rotating black hole.
Indeed, surfaces at $g_{tt}=0$, in the region outside the external
horizon fill the interval $[r_m, r_e]$ where $r_+ < r_m = (2m\ell^2)^{1/3}$
and $r_e$ is the
positive root of $r^4 -2m\ell^2r-\lambda^4a^4=0$. 
(There is another surface at
$g_{tt}=0$ for
$r<0$, filling the interval $[r'_e,0]$ where $r'_e$ is the remaining negative
solution of the above equation.)

\subsection{Mass and angular momentum}

With the 
given choice of the periods ($T$ for $P$ and $2\pi$ for $\phi$), 
we determine the area of the event horizon to be
\begin{eqnarray}
 {\cal A}=2\pi Tr_+^2
\end{eqnarray}
and the angular velocity, $\Omega_H=ar_+^{-2}$. The Hamiltonian mass of 
the given spacetime, relative to the background solution with toroidal
topology but $m=a=0$, can be computed by carefully handling the 
divergent terms appearing when the boundary of spacetime is pushed to 
spatial infinity. Also, the Killing observers at infinity relative to 
which the mass is measured, have a residual, $P$-dependent angular 
velocity
\begin{eqnarray}
\Omega_{\infty}=a\ell^{-2}P^2
\end{eqnarray}
and this also must be taken into account. All calculations done, we 
get a conserved mass $M$ and a conserved angular momentum $J$
according to
\begin{eqnarray}
M=\frac{mT}{2\pi}, \qquad J=Ma,
\end{eqnarray}
giving to $a$ the expected meaning. As $a$ is bounded from above by 
$a_c$, in order for the solution to have a black hole interpretation, 
we see that the angular momentum is bounded by a power $M^{5/3}$ of 
the mass. 

\section{Conclusion}

We have presented a class of exact solutions of Einstein's equations 
with negative cosmological constant, having many of the features which 
are characteristics of black holes. All these solutions are of
Petrov-type $D$ and the horizons, when they exist, have the topology 
of Riemann surfaces and therefore they lack rotational symmetry for 
genus $g>1$. Among the solutions there is also a 
toroidal black hole, still different from the Lemos-Zanchin
solution. The toroidal metric 
has an exact rotational symmetry and a well-defined 
mass and angular momentum. From this perspective, it is more promising as 
a thermodynamical object and one may hope to find suitable 
generalizations of the "four laws of black hole 
mechanics" \cite{habaca}. Apart from this, the solutions seem to be 
interesting in their own rights, they have intriguing properties 
and may provide further ground to test string theory ideas 
in black hole physics and the character of singularities in general 
relativity.

\section*{Acknowledgement}

The part of this work due to D.K. has been
supported by a research grant within the
scope of the {\em Common Special Academic Program III} of the
Federal Republic of Germany and its Federal States, mediated 
by the DAAD.\\ 
The part of this work due to V.M. has been financially supported
by the ECT* (European Centre for Theoretical Studies in Nuclear
Physics and Related Areas) of Trento, Italy.

\end{document}